\documentclass[conference]{IEEEtran}%{sig-alternate-05-2015}

\usepackage{amssymb,amsmath,url,rotating,ifthen,algorithm,algorithmicx,epsfig,multirow,array,color,multicol,graphicx,caption,subcaption,float,hhline,calc,enumitem}
\usepackage{todonotes} 
\usepackage{bold-extra}

\usepackage{algpseudocode}% http://ctan.org/pkg/algorithmicx
\algnewcommand\algorithmicinput{\textbf{Input:}}
\algnewcommand\INPUT{\item[\algorithmicinput]}
\algnewcommand\algorithmicoutput{\textbf{Output:}}
\algnewcommand\OUTPUT{\item[\algorithmicoutput]}

\algrenewcommand{\algorithmicforall}{\textbf{for each}}

\def\NN{\mathcal{N}}

\def\e{\epsilon }
\def\chi{{\mathbf 1}}

\def\R{{\mathbb R}}

\begin{document}
\title{Surprising strategies obtained by stochastic optimization in partially observable games}
%	\institute{ TAO (Inria), LRI, UMR 8623 (CNRS - Univ. Paris-Sud)}
%\author{Marie-Liesse Cauwet\inst{1}, Olivier Teytaud\inst{1}}
%\numberofauthors{4}	

% NOT double-blind!
\author{
\IEEEauthorblockN{ Marie-Liesse Cauwet}
\IEEEauthorblockA{\textit{\small Mines Saint-Etienne, Univ Clermont Auvergne,}\\
\textit{\small CNRS, UMR 6158 LIMOS, Institut Henri Fayol, Departement GMI,}\\
\small F - 42023 Saint-Etienne France\\
\small marie-liesse.cauwet@emse.fr}
\and
\IEEEauthorblockN{Olivier Teytaud}
\IEEEauthorblockA{\textit{\small TAO} \\
\textit{\small Inria Saclay}\\
\small Gif-Sur-Yvette, France \\
\small olivier.teytaud@inria.fr}
}

\maketitle

\begin{abstract}
This paper studies the optimization of strategies in the context of possibly randomized two players zero-sum games with incomplete information. We compare $5$ algorithms for tuning the parameters of strategies over a benchmark of $12$ games. A first evolutionary approach consists in designing a highly randomized opponent (called naive opponent) and optimizing the parametric strategy against it; a second one is optimizing iteratively the strategy, i.e. constructing a sequence of strategies starting from the naive one. $2$ versions of coevolutions, real and approximate, are also tested as well as a seed method. The coevolution methods were performing well, but results were not stable from one game to another. In spite of its simplicity, the seed method, which can be seen as an extremal version of coevolution, works even when nothing else works. Incidentally, these methods brought out some unexpected strategies for some games, such as Batawaf or the game of War, which seem, at first view, purely random games without any structured actions possible for the players or Guess Who, where a dichotomy between the characters seems to be the most reasonable strategy. 
%We consider the noisy optimization of parametric policies in partially observable games. We compare 5 optimization strategies: a naive approach based on designing a highly randomized opponent and optimizing the parametric policy in front of it, an iterative one, the seed method, and 2 versions of coevolution. In spite of theoretical shortcomings, the iterative method and the approximate coevolution method appear to be reliable; in spite of its simplicity, the seed method works even when nothing else works. We then study in details some unexpected strategies found by these methods for some specific games.  
%In spite of theoretical shortcomings, a simple approach based on (i) designing a highly randomized opponent (ii) optimizing a parametric policy in front of it, proves to be a reliable approach.
All source codes of games are written in Matlab/Octave and are freely available for download.
\end{abstract}

%\category{G.1.6}{Optimization}{Unconstrained optimization}

%\terms{Theory}

%\keywords{Noisy optimization}
%\setcounter{tocdepth}{4}
%\tableofcontents
%\setcounter{tocdepth}{4}

\section{Introduction}

In game theory, the case of adversarial problems where two agents try to optimize antagonist rewards under incomplete information is an important class of problems with real-life applications in games, in robust optimization and in military applications such as mission planning~\cite{tristanMilitaryPaper}. 

In a game, if the number of deterministic strategies (called pure strategies) is finite, some stochastic strategies (called mixed strategies or arms) are usually represented by vectors of non-negative numbers summing to $1$: a player following a stochastic strategy $x=(x_1,\dots,x_N)$ will adopt the $i^{th}$ deterministic strategy with probability $x_i$. Defining optimality is delicate in such a setting. We might have $x$ performing better than $x'$ against $y$, but $x'$ performing better than $x$ against $y'$.
A classical approach is to find an optimum in the Nash sense~\cite{vonneumann,nash52}. In the case of finite and moderate number of pure strategies, the literature proposes some efficient solutions \cite{lpnash,grigo95sublinear,Auer02}. 

%Consider an adversarial problem: two agents try to optimize antagonists rewards.
%This situation occurs naturally:
%	in games;
%	in robust optimization;
%	in military applications, such as mission planning~\cite{tristanMilitaryPaper}.
%When the set of (deterministic) policies is finite, (stochastic) policies are usually represented by vectors of non-negative numbers summing to 1; with (stochastic) policy $x$, the $i^{th}$ deterministic policy is chosen with probability $x_i$. Stochastic policies are termed mixed policies, whereas the deterministic policies are termed pure policies, or arms. Defining optimality is tricky in such a setting; we might have $x$ performing better than $x'$ against $y$, but $x'$ performing better than $x$ against $y'$.
%A classical approach is to find an optimum in the Nash sense~\cite{vonneumann,nash52}. Given a function $f$, if the first agent chooses a mixed policy $x$ and the second agent chooses a mixed policy $y$, then the reward for the first agent is $f(x,y)$ and the reward for the second agent is $-f(x,y)$. A Nash equilibrium is then a pair $(x^*,y^*)$, such that for all $x$, $f(x,y^*)\leq f(x^*,y^*)$ and for all $y$, $f(x^*,y)\geq f(x^*,y^*)$. It is known that, if the state of pure policies is finite, $(x^*,y^*)$ is not necessarily unique, but $f(x^*,y^*)$ is unique. There are various algorithms for solving such problems in the finite case %most of them focusing on moderate sizes~
%\cite{lpnash,grigo95sublinear,Auer02}. 

We are focusing on real world partially observable games for which computing an approximate Nash equilibrium requires heavy computations \cite{mundhenk,rintanen,auger2011} and computing an exact Nash equilibrium is undecidable~\cite{auger2012frontier,saffidine:hal-01256660} when the horizon is unbounded. Indeed, there are infinitely many pure strategies, so that the existence of a Nash equilibrium is not guaranteed.

In this framework, we propose a simpler but more practical approach: we will consider some parametric strategies. If the parameters take a finite number of values, this boils down to the classical multi-armed bandits problem~\cite{manyarmed08}. Here, we consider some parameters taking values in an interval of $\R$ (hence infinitely many arms) and optimize them against a baseline~\cite{ea2015paperongames}.

%Hence, we consider the case of a parametric policy. Optimizing this policy using approaches above is then a large scale problem, requiring infinitely many arms~\cite{manyarmed08}. However, when the policies are clearly ranked (as in the Elo model~\cite{elo}), the structure is quite modified; mixed Nash equilibria are not necessary and simpler approaches might be more practical. As pointed out in \cite{ea2015paperongames}, simply optimizing against a baseline might be a good approach.

%We here compare (i) a simple approach (design a very randomized baseline, covering ``approximately equally'' all the state space) and optimize a parametric policy (ii) a more sophisticated approach, optimizing iteratively a policy $\pi_{n+1}$ against $\pi_n$; after disappointing preliminary results.%, we discarded coevoution; a detailed comparison is left as further work.% (iii) coevolution.

We implemented 5 methods to optimize a parametric strategy: a naive evolutionary algorithm, a more sophisticated (iterative) one, 2 variants of coevolution and a seed method. 
The naive approach consists in first designing a very randomized baseline, covering ``approximately equally'' all the state space and then  optimizing a parametric strategy against it. The more elaborate approach is to optimize iteratively a strategy $\pi_{n+1}$ against $\pi_n$. Coevolution is a different approach~\cite{coevolbackgammon,cocmaes}. A population $P_1$ of agents evolves for the role of agent 1, and another population $P_2$ evolves for the role of agent 2. Population $P_1$ is optimized against population $P_2$, whereas population $P_2$ is optimized against population $P_1$. The seed method is an extreme case of coevolution without crossover or mutation (just random generation and selection at the end) - generate a population as large as you can for statistically analyzing it and just pick up the best. 
Section~\ref{sec:algo} describes these methods.

%We conclude that the first approach above is often, in spite of theoretical shortcomings, an excellent and robust approach; that the second one might be excellent sometimes, but is unstable in many cases.%; and that the third, most principled, approach is often unrealistic. 

%We get surprisingly strong policies at games where it is not that obvious that strong policies exist: 
%\begin{itemize}
%\item Batawaf and War (card game of war) for which most people believe that there is no policy; 
%\item Guess who (for which most people believe that Dichotomy is the best policy);
%\item Battleship, where it is not that obvious that there is something beyond the Monte Carlo estimation of hitting probabilities.
%\item Cheat and Flip, where TODO
%\end{itemize}

We first compare these methods on various games (presented in Section \ref{games}) in Section~\ref{sec:comp}. Then we use the obtained optima in Section~\ref{sec:detailed} to determine some specific, detailed strategies in games where it is not so obvious that strong strategies exist (a strategy is strong if the probability of winning - when using it against a fixed set of other strategies - is significantly higher than $50$\%):
\begin{itemize}
\item Batawaf and War (card games) for which most people believe that there is no clever strategy; 
\item Guess who, for which most people believe that dichotomy is the best strategy;
\item Battleship, where it is not so obvious that there is something beyond the Monte-Carlo estimation of hitting probabilities.
\end{itemize}

\section{Algorithms}\label{sec:algo}

\subsection{Preliminaries: Bernstein race as a comparison operator}

Let us assume that we wish to compare two players, $p_a$ and $p_b$.
There are several possibilities.

In the simplest method, we play $101$ games between $p_a$ and $p_b$ and select the one which wins at least 51 games. Selection can go wrong: if the two players have close levels, the weakest player can be selected, with probability arbitrarily close to 50\%.

Let us therefore introduce a statistical test.
A simple statistical test consists in playing games between $p_a$ and $p_b$, and perform statistical tests until one of them has won significantly more games than the other. A statistical test has a risk of failure $\delta$. If $20$ tests are performed, then the risk that at least one test fails is $1-(1-\delta)^{20}\approx 20\delta$ for $\delta$ small enough (by Taylor expansion). Therefore, without any correction, the naive statistical test method might fail with probability arbitrarily close to 50\%.

An alternative is to play games between $p_a$ and $p_b$, and perform statistical tests until one of them has won significantly more games than the other, taking into account the Bonferroni correction~\cite{bonferroni36} or other tests taking into account the multiple nature of these tests. This is the \textit{Bernstein race}~\cite{Mnih08}. Races take care of correcting the statistical test, in order to guarantee a given error rate $\delta$. Whatever are the two players, the probability that the weakest is selected is at most $\delta$. A drawback is that it might take a lot of time, and if the two players have exactly the same level (in the sense that the probability of winning is 50\%), then there is a probability $1-\delta$ that the race never halts. We then apply a \textit{limited race}: this is the same as for Bernstein races, except that when the winning rate is known up to a given precision ($0.01$ in our experiments), the race is stopped and the winner is the current best.

%\begin{itemize}
%	\item Naive method: Play 101 games between P1 and P2; select the one which wins at least 51 games.
%	\item Naive statistical test: Play games between P1 and P2, and perform statistical tests until one of them has won significantly more games than the other.
%	\item Race: Play games between P1 and P2, and perform statistical tests until one of them has won significantly more games than the other. Take into account the Bonferroni correction.
%	\item Limited race: the same as for races, except that when the winning rate is known up to a given precision, the race is stopped and the winner is the current best.
%\end{itemize}

% The Bonferroni correction\cite{bonferroni} is as follows: a statistical test has a risk of failure, possibly denoted $\delta$; if you perform 20 tests, the risk of failure is $20\delta$. Therefore, without Bonferroni correction (i.e. the naive statistical test method above) might fail with probability arbitrarily close to 50\%.

% Races take care of correcting the statistical test, in order to guarantee a given error rate $\delta$. Whatever maybe the two players, the probability that the weakest is selected is at most $\delta$. A drawback is that it might take a lot of time, and if the two players have exactly the same level (in the sense that the probability of winning is 50\%), then there is a probability $1-\delta$ that the race never halts. We refer to \cite{bernsteinStopping} for more on Bernstein races. We here apply a limited race, with target precision $0.01$.

\subsection{Naive evolutionary algorithm}
A first method, termed naive (see Alg. \ref{algoponaive}) consists in simply optimizing the average performance against a baseline, for example the default strategy using random independent standard Gaussian parameters. 

\begin{algorithm}
{\footnotesize
\begin{algorithmic}
\State {\bf Input}: a game simulator, a precision parameter $\e$, parameters of the random opponent $x^0$
\State $\sigma \leftarrow 1$ \Comment{Initial step-size}
\State $x\gets x ^0$\Comment{Initial strategy}
\While{ (termination criterion is not met)}
	\For{$i = 1$ to length of $x$}
	\State $x'_i\leftarrow x_i+ \sigma \NN(0,1)$ \Comment{Componentwise Mutation}
    \EndFor
	\Repeat
        \State{ play a game:}
		\begin{itemize}
			\item between $x$ and $x^0$
			\item between $x'$ and $x^0$
		\end{itemize}
	\Until{the limited Bernstein race of precision $\e$ stops}
	\If{$x'$ performed better than $x$ against $x^0$}
	\State $x\leftarrow x'$
	\State $\sigma\leftarrow 2\sigma$
	\Else
	\State $\sigma\leftarrow 0.84\sigma$
	\EndIf
\EndWhile
\State {\bf Output}: an approximation $x$ of the optimal strategy
\end{algorithmic}
}
\caption{\label{algoponaive}\small Naive algorithm. $\NN(a,b)$ denotes a Gaussian variable of mean $a$ and standard deviation $b$. $x_i$ denotes the $i^{th}$ component of $x$.}
\end{algorithm}

\subsection{Iterative evolutionary algorithm}
A second simple intuitive algorithm is to accept a search point as a new baseline as soon as it is statistically better (winning rate $>50\%$) than the previous one. Many evolutionary algorithms are comparison-based and can be adapted easily to that framework. We can for example apply the $(1+1)$ Evolution Strategy\footnote{Note that here, and only here, the denomination ``strategy'' refers to a type of evolutionary algorithm. In the rest of the paper, ``strategy'' denotes the player game plan, hence there is no ambiguity.} \cite{rech} with games as a comparison operator, as proposed in Alg.~\ref{algopo}.

\begin{algorithm}
{\footnotesize
\begin{algorithmic}
\State {\bf Input}: a game simulator and a precision parameter $\e$, parameters of the initial opponent $x$
\State $\sigma \leftarrow 1$ \Comment{Initial step-size}
%\State In case of the Parisian or Robust variant, initialize the archive at the singleton $x$.
\While{ (termination criterion is not met)} 
	\For{$i=1$ to length $x$}
	\State $x_i'\leftarrow x_i+ \sigma\NN(0,1)$ \Comment{Componentwise Mutation}
    \EndFor
	\Repeat
    \State { play games between $x'$ and $x$ }
	\Until{the limited Bernstein race of precision $\e$ stops}
%	\State{Robust variant: do the same comparison for all other $x$ in the archive}
	\If{ $x'$ better than $x$}% (Robust case: all $x$ in the archive)}
	\State $x\leftarrow x'$
%	\State In case of Parisian or Robust variant, store $x'$ in the archive.
	\State $\sigma\leftarrow 2\sigma$
	\Else
	\State $\sigma\leftarrow 0.84\sigma$
	\EndIf
\EndWhile
\State {\bf Output}: an approximation $x$ of the optimal strategy
%\State PI: Output the policy randomly choosing one of the element in the archive as an approximation of the optimum.
\end{algorithmic}
}
\caption{\label{algopo}\small Iterative approach. $\NN(a,b)$ denotes a Gaussian variable of mean $a$ and standard deviation $b$. $x_i$ denotes the $i^{th}$ component of $x$.}%; the Parisian Iterative variant (PI) consists in using as an output the uniform distribution over the archive. In the Robust variant (RI), a mutation is validated if and only if outperforms all players in the archive.}
\end{algorithm}

At first view, this algorithm looks better than the naive one. We compare against a stronger opponent, so winning rates should be more informative.
A key question is however whether we might have a \textit{red queen effect}: algorithms winning against the previous generation, but becoming very specialized, and unable to compete with even simple tools~\cite{kauffman1995escaping,vanvalen1973}.
%We might also want to consider a Parisian version of this algorithm; this means that we use the whole population for generating a solution; and, by analogy with classical works in coevolution, we will define our output by a random policy, uniformly drawn among the population, each time a game is played. 

\subsection{Coevolution}

A typical coevolution considers Black players and White players (in case of a game with a Black player and a White player), and the fitness of the Black player is the average score against the White population, while the fitness of a White player is the average score against the Black population. A classical technique (e.g. applied in \cite{grigo95sublinear}) considers ``extended'', ``doubled'' players, who have two parts: one for playing as Black, and one for playing as White. Then, we need only one population. This is applied in the present paper.

In the first coevolution variant, called \textit{real coevolution}, a search point is accepted in the population if it is statistically better than {\textit{every}} point of this population. It is presented in Alg.~\ref{alg:coevol}.

\begin{algorithm}
{\footnotesize
\begin{algorithmic}
\State {\bf Input}: a game simulator and a precision parameter $\e$, parameters of the initial opponent $x$.
\State $\sigma \leftarrow 1$ \Comment{Initial step-size}
\State $P\gets \{x\}$ \Comment{Best points population}
\While{ (termination criterion is not met)}
	\For{$i =1$ to length of $x$}
	\State $x_i'\leftarrow x_i+ \sigma\NN(0,1)$\Comment{Mutation}
    \EndFor
	\Repeat
    \State {~play a game between $x'$ and every point of $P$}
	\Until{~each limited Bernstein race of precision $\e$ stops}
    \If{$x'$ performed better than all points in $P$}
	\State $x\leftarrow x'$
	\State $P\gets \{P,x'\}$
	\State $\sigma\leftarrow 2\sigma$
	\Else
	\State $\sigma\leftarrow 0.84\sigma$
	\EndIf
\EndWhile
\State {\bf Output}: an approximation $x$ of the optimal strategy
\end{algorithmic}
}
\caption{\label{alg:coevol} \small Real coevolution. $\NN(a,b)$ denotes a Gaussian variable of mean $a$ and standard deviation $b$.  $x_i$ denotes the $i^{th}$ component of $x$.}
\end{algorithm}  
%A population $P_1$ maximises $x\mapsto f(x,y)$ on average over $y\in P_2$ whereas $P_2$ minimizes $y\mapsto f(x,y)$ on average over $x\in P_1$.

We might consider the population as a representation of a stochastic strategy, i.e., the global strategy is randomly drawn in the population for each game the artificial intelligence had to play: this is the so-called ``Parisian approach'' ~\cite{ParisianApproach}. Our second variant of coevolution, called \textit{approximate coevolution} (see Alg.~\ref{alg:approx}), is related to this approach. Instead of comparing the search point to each point of the population, it is only compared to one point of the population drawn at random.

\begin{algorithm}
\footnotesize
{\begin{algorithmic}
\State {\bf Input}: a game simulator and a precision parameter $\e$, parameters of the initial opponent $x$. 
\State $\sigma \leftarrow 1$ \Comment{Initial step-size}
\State $P\gets \{x\}$ \Comment{Best points population}
\While{ (termination criterion is not met)}
	\For{$i=1$ to length of $x$}
	\State $x'_i\leftarrow x_i+ \sigma \NN(0,1)$ \Comment{Componentwise Mutation}
    \EndFor
	\State Draw at random an integer $rand$ between $1$ and the size of $P$
	\Repeat{ play a game between $x'$ and the $rand^{th}$ individual of $P$}
	\Until{the limited Bernstein race of precision $\e$ stops}
	\If{$x'$ performed better}
	\State $x\leftarrow x'$
	\State $P\gets \{P,x'\}$
	\State $\sigma\leftarrow 2\sigma$
	\Else
	\State $\sigma\leftarrow 0.84\sigma$
	\EndIf
\EndWhile
\State {\bf Output}: an approximation $x$ of the optimal strategy
\end{algorithmic}
}
\caption{\label{alg:approx}\small Approximate coevolution. $\NN(a,b)$ denotes a Gaussian variable of mean $a$ and standard deviation $b$. $x_i$ denotes the $i^{th}$ component of $x$.}
\end{algorithm}

\subsection{Method of seeds}
\cite{autoreinforcegameai,rectangular,poseed,fastseed,ictai2017} proposed a new method for optimizing an artificial intelligence, which can be applied to our setting.
The principle is to generate a population of $K$ random individuals for the first player, and another population of $K$ random individuals for the second player. Then, we consider all pairwise games
(a more subtle method was proposed in \cite{fastseed} but we do not consider this in the present paper).
We then select the player with best average performance. Both populations can be equal when the game is symmetric, and we might increase $K$ until the time budget is elapsed in order to be anytime.

\section{Games}\label{games}
The games are fully described in our open source platform (https://gforge.inria.fr/projects/gametestbed/). We provide a short description below. 

\subsection{Brief description}

\paragraph*{Batawaf and game of War} The cards, kept unknown from the two players, are randomly distributed among them. At each turn, each player reveal simultaneously the card at the top of their deck. The player with the highest valued card wins the cards and put them at the bottom of his/her deck. In case of draw, the two players place a second card, face down, on the pile, then a third one face up. The process is possibly repeated until the winner of the turn can be determined. The player who gets all the cards wins the game. The difference between Batawaf and War is the total number of cards and in the number of cards of same strength.

\paragraph*{ Battleship} Two players place secretly four ships of size $2$, $3$, $4$ and $5$ on a $9\times 9$ board without crossing, horizontally or vertically. In each round, each player in turn propose a position on the board, trying to find out a ship of the adversary. The first player to have found all parts of all ships of her/his opponent wins. At each turn the player is informed of whether he has touched something.

\paragraph*{Cheat}
The cheat game uses 52 cards of 4 different colors. The cards are randomly distributed among the players. One of the players puts a first card on the table. The color of this card is termed the ``current color''. Then, each player, in turn, either puts a card (face hidden) on top of the previous one\footnote{(s)he can choose the color of the card.} or claims ``cheat''. In the latter case, if the last played card was of the current color, the person who claimed ``cheat'' receives all the cards currently on the table. Otherwise, the previous player receives all the cards currently on the table. In both cases, a new round start, with the player who received all the cards putting a chosen card, which decides the new current color. The first player who gets rid of all his/her cards wins the game.
%This game is widely played in France, Chile, TODO
%The cheat game can also be played with $N>2$ players.

\paragraph*{Flip} In the flip game, cards are randomly distributed among the two players. Only the five topmost cards are visible (for both players). When a card is removed, another one is made visible.
Both players, in turn\footnote{The game is also played as a speed game, with players playing as soon as they can instead of playing in turn.}, put one of their cards (if they can) on top of one of the two visible cards on the table. A card may be put on top of the card if it is the card just above or just below in the ranking - except that the strongest card can be put on the weakest and the weakest can be put on the strongest. 
The first player who gets rid of all his cards wins the game.

\paragraph*{Guess who} Each player privately draws a number between $1$ and $128$. At each round, in turn, each player asks the other if his/her number is in the first $x$ numbers (the player chooses $x$). The answer must be sincere, so that some possibilities can be can removed. The first player who determines the number chosen by the other wins.

%\paragraph{Nim}
% The Nim game is well known; 4 rows of sticks are installed, with 1, 3, 5 and 7 sticks respectively. Each player, in turn, must remove 1, 2 or 3 sticks, on a same row (chosen by himself/herself). The player who removes the last stick loses the game.
%The Nim game is also known as TODO, it is played in TODO.
%The optimal policy can be computed analytically. {\color{magenta}{Is it possible to compare to strategy obtained by simulation is similar to the optimale theoretical one ? how different is it ? can we play a game between these 2 strategies ?}}

\paragraph*{Morra} In the Morra game, both players simultaneously announce (i) a number of fingers (usually by showing with the hand) between 0 and 5 and (ii) a guessed number between 0 and 10. If the sum of the numbers of fingers is equal to the guess of one (and only one) player, then this player wins the game (otherwise, the game is a draw).

\paragraph*{Phantom 4 in row} It is the stochastic version of 4 in row. In a 4x7 grid, at each turn, the players choose a permutation of the 7 locations and a move, without observation of the game. If the chosen location is already full, the stone goes to the next location according to the permutation. The first player with 4 stones aligned wins.

\paragraph*{Phantom tic tac toe} The tic tac toe game, in 3x3, is well known and not very challenging. In the phantom version, a permutation of the 9 locations is chosen. A player cannot see her opponent's moves. Whenever a player chooses a location already occupied, the move is switched to the next location in the order of the permutation.
Finding a strong strategy in Phantom tic tac toe is far less trivial than for tic tac toe.

\paragraph*{Other games} Other games in the platform are not partially observable: Pig is a dice game, Nim (a.k.a. Marienbad game) is well known as a misery game; four-in-a-row is famous. For these games our approaches are not likely to be meaningful and these results are just included for shedding light on the genericness of our methods.

\subsection{Parametric strategies}
The strategies will not be detailed here. They can be downloaded at http://www.lri.fr/$\sim$teytaud/gametestbed.html. They are roughly described in Table \ref{tabpolicies}.
\begin{table*}[!ht]
\centering
\footnotesize
% RC=opocoevol (>=50% against all previous)
% C=opovcoevol (>=50% on average against archive)
% I=opo (>=50% against previous)
% N=oponaive (>= previous in terms of winning rate against naive)
\scriptsize
{\setlength{\tabcolsep}{0.3em}
\begin{tabular}{|p{3cm}|c|p{7.5cm}|c|c|}
\hline
Game & Number of & Info & Stochasticy of the & The optimal strategy \\
     & parameters &     & strategies & must be stochastic\\     
\hline
War & 3/4 & Probability of selecting the ascending order, prob. of selecting the
descending order, probability of random order, perturbation of the selected ranking, see Section \ref{bw} & No/Yes     & ?          \\
\hline
Batawaf & 3/4 & idem War & idem  & idem \\     
\hline
Battleship          & 33  & Parameters of the probability distribution for the ships & Yes & Yes \\
\hline
Cheat                & 28/29  & Parameters of a randomly perturbed quadratic value function;
the $29^{th}$ optional parameter (variant called ``special'') is an arbitrary perturbation of the value function, built pseudo-randomly from that parameter & Yes & Yes \\
\hline
Flip                & 420 & Parameters of a quadratic value function & No & No\\
\hline
Guess who (deterministic/stochastic)          & 4/5 & Nonlinear expert strategy based on a risk level based on the gap (see \cite{guesswho} and Section \ref{gw}) & No/Yes & No \\
\hline
Morra               & 66  & Vector of probabilities for all possible joint actions & Yes & Yes \\
\hline
Nim                 & 383 & Complete value function over the 383 non-terminal states& No & No \\
\hline
Phantom 4-in-a-row              & 2 & Complete pure strategy (i.e. index of the chosen list of moves when playing first and index of the chosen list of moves when playing second) & No & ? \\
\hline	 	      
Phantom tic-tac-toe            & 18 & Complete pure strategy& No & Yes\\
\hline
4-in-a-row       & 14 & Parameter of the Monte-Carlo strategy & Yes$^{\ast}$& No \\
\hline
\end{tabular}
}
\caption{\label{tabpolicies}\small Characteristics of the parametric strategies for each game: number of parameters and stochasticity or not of the strategies. $^{\ast}$ denote ``yes'' for strategies which are stochastic only by the finiteness of samples and/or draws in Monte-Carlo simulations. %The best method (if any) if the one which wins with probability at least $50\%$ against all others (for ``4 in a row", it was the case for two methods; naive performed best TODO 100\% is possible).  
}
\end{table*}

\section{Results}\label{sec:comp}

\paragraph*{Scores against the baseline} Table \ref{poula} presents the score of each method against the original default parametrization. Each learning is performed only once; methods in which no iteration was obtained are not displayed. Each method was run 24 hours. Seed was not run for ``Guess Who'' variants.

\begin{table*}
\centering
\scriptsize
{\setlength{\tabcolsep}{0.3em}
\begin{tabular}{|c|c|c|}
\hline Game & Method & Score against baseline \\
\hline
\multirow{4}{*}{4-in-a-row} & Seed & 0.63 $\pm$ 0.03              \\
&{̀\bf Iterative }& 1 $\pm$ 0              \\
&{̀Naive }& 0.75 $\pm$ 0.19        \\
&{̀Real coevol }& 0.442308 $\pm$ 0.10           \\
\hline
\multirow{5}{*}{Batawaf4} & Seed & 0.606 $\pm$ 0.004 \\
& Coevol & 0.626 $\pm$ 0.005\\
&{̀\bf  Iterative }& 0.626 $\pm$ 0.004\\
&{̀Naive }& 0.620 $\pm$ 0.004\\
&{̀Real coevol }& 0.544 $\pm$ 0.003 \\
\hline
\multirow{5}{*}{Batawaf} & Seed  & 0.627 $\pm$ 0.004 \\
&Coevol & 0.626912 $\pm$ 0.005\\
 &Iterative & 0.626671 $\pm$ 0.004\\
&{̀\bf Naive }& 0.628467 $\pm$ 0.004\\
&{̀Real coevol }& 0.443975 $\pm$ 0.003\\
\hline
\multirow{5}{*}{Cheat} &{\bf Seed} & 0.845 $\pm$ 0.003 \\
&{̀cheat, Coevol }& 0.783318 $\pm$ 0.006\\
&{̀ Iterative }& 0.817416 $\pm$ 0.029\\
&{̀ Naive }& 0.802905 $\pm$ 0.011\\
&{̀Real coevol }& 0.810882 $\pm$ 0.007\\
\hline
\multirow{5}{*}{Flip} &{ Seed} & 0.554 $\pm$ 0.04 \\
&{̀\bf Coevol }& 0.597727 $\pm$ 0.017\\
&{̀ Iterative }& 0.588 $\pm$ 0.022  \\
&{̀ Naive }& 0.593143 $\pm$ 0.017\\
&{̀ Real coevol }& 0.591314 $\pm$ 0.016\\
\hline
\end{tabular}
\begin{tabular}{|c|c|c|}
\hline 
Game & Method & Score against baseline \\
\hline
\multirow{3}{*}{Guess who-deter} &{̀\bf  Coevol }& 0.785714 $\pm$ 0.113804\\
&{̀Naive }& 0.613484 $\pm$ 0.000513042 \\
&{̀Real coevol }& 0.597514 $\pm$ 0.00160714  \\
\hline
\multirow{4}{*}{Guess who} &{̀Coevol }& 0.679$\pm$ 0.130\\
&{̀ Iterative }& 0.771$\pm$ 0.0876 \\
&{̀\bf  Naive }& 1 $\pm$ 0    \\
&{̀ Real coevol }& 0.615 $\pm$ 0.00158 \\
\hline
\multirow{5}{*}{Cheat-Special} &{ Seed} & 0.6405 $\pm$ 0.02 \\
&{̀Coevol }& 0.64 $\pm$ 0.016\\
&{̀ Iterative }& 0.58 $\pm$ 0.01\\
&{̀\bf  Naive }& 0.65 $\pm$ 0.027\\
&{̀ Real coevol }& 0.62 $\pm$ 0.027\\
\hline
Battleship & {\bf Seed} & 0.77 \\
\hline
\multirow{5}{*}{Morra}& { Seed} & 0.519 $\pm$ 0.0013 \\
&{̀ Coevol }& 0.526 $\pm$ 0.0012 \\
&{̀\bf  Iterative }& 0.527 $\pm$ 0.0016 \\
&{̀ Naive }& 0.523$\pm$ 0.001 \\
&{̀ Real coevol }& 0.518 $\pm$ 0.001 \\
\hline
\multirow{4}{*}{Nim} & {\bf Seed} & 0.74 $\pm$ 0.0003 \\
&{̀ Coevol }& 0.500 $\pm$ 0.0007\\
&{̀ Iterative }& 0.589 $\pm$ 0.002\\
&{̀ Real coevol }& 0.669 $\pm$ 0.002\\
\hline
\end{tabular}
\begin{tabular}{|c|c|c|}
\hline 
Game & Method & Score against baseline \\
\hline
\multirow{4}{*}{Phantom-4-in-a-row}&{\bf Seed } & 0.69 $\pm$ 0.003\\
&{̀ Coevol }& 0.549$\pm$ 0.00096\\
&{̀ Iterative }& 0.657$\pm$ 0.004\\
&{̀ Real coevol }& 0.598 $\pm$ 0.003\\
\hline
\multirow{5}{*}{Phantom-tic-tac-toe} & {\bf Seed } & 0.75 $\pm$ 0.0012 \\
&{̀ Coevol }& 0.445 $\pm$ 0.000497\\
&{̀ Iterative }& 0.595 $\pm$ 0.00138\\
&{̀ Naive }& 0.445 $\pm$ 0.000497\\
&{̀ Real coevol }& 0.696 $\pm$ 0.0011\\
\hline
\multirow{5}{*}{Pig} & {Seed } & 0.505 $\pm$ 0.003 \\
&{Coevol }& 0.507889 $\pm$ 0.0012\\
&{̀\bf  Iterative }& 0.522151 $\pm$ 0.002\\
&{̀ Naive }& 0.512262 $\pm$ 0.0016 \\
&{̀ Real coevol }& 0.508092 $\pm$ 0.0016\\
\hline
\multirow{5}{*}{War4} & {Seed} & 0.623 $\pm$ 0.017 \\
&{̀ Coevol }& 0.623288 $\pm$ 0.016\\
&{̀\bf  Iterative }& 0.669$\pm$ 0.05\\
&{̀ Naive }& 0.605$\pm$ 0.018\\
&{̀ Real coevol }& 0.553$\pm$ 0.010\\
\hline
\multirow{5}{*}{War} &{ Seed} & 0.632 $\pm$ 0.003 \\
&{̀ Coevol }& 0.517 $\pm$ 0.00666\\
&{̀\bf Iterative }& 0.653 $\pm$ 0.012\\
&{̀ Naive }& 0.632 $\pm$ 0.018\\
&{̀ Real coevol }& 0.487  $\pm$ 0.005\\
\hline
\end{tabular}
}
\caption{\label{poula}\small Scores (winning rates) of each method against the original default parametrization. The standard deviation is shown after $\pm$. Despite the fact that the naive method is directly optimizing the target criterion, other methods are competitive. The seed method is particularly robust. Batawaf4 (resp. War4) denotes the Batawaf (resp. War) strategy with $4$ parameters.
}
\end{table*}

\paragraph*{Comparison between 4 optimization methods}
We now compare strategies obtained by our different methods against each other rather than against the original baseline.
Numerical results are displayed in Table \ref{popof}. %the iterative method sometimes fails (in the sense that it performs poorly either against the original baseline or against the Naive method). % - it obtains quite good results for the only fully observable game, namely Nim, which is not the goal of the present work (Nim can easily be solved exactly). Detailed results at Guess Who (Section \ref{gw}) suggest that this is a red queen effect.
All results are obtained after 24 hours of learning.% the Iterative method was somehow unstable compared to the simple Naive method. {\color{magenta} apart for Cheat, it's always better or similar to naive method.}
\begin{table*}[!ht]
\begin{subtable}{0.5\linewidth}
\centering
\scriptsize
{\setlength{\tabcolsep}{0.3em}
\begin{tabular}{|c|c|c|c|c|}
\hline
\multicolumn{5}{|c|}{Cheat (28 params)}\\
\hline
 & vs naive & vs itera. & vs ap. coevol & vs coevol \\
\hline
naive& 0.5(+-  0)	& 0.0269(+-0.003)	& 0.017(+-0.002)	& 0.977(+-0.002)	\\
iterative& 0.973(+-0.003)	& 0.5(+-  0)	& 0.5(+-0.008)	& 0.502(+-0.008)	\\
{\bf{ approx. coevol}}& 0.983(+-0.002)	& 0.5(+-0.008)	& 0.5(+-  0)	& 0.671(+-0.008)	\\
real coevol& 0.0235(+-0.002)	& 0.498(+-0.008)	& 0.329(+-0.008)	& 0.5(+-  0)	\\
\hline
\multicolumn{5}{|c|}{GuessWho-deter (4 params)}\\
\hline
 & vs naive & vs itera. & vs ap. coevol & vs coevol \\
\hline
naive& 0.5(+-  0)	& 0.8(+-0.004)	& 0.489(+-0.005)	& 0.498(+-0.005)	\\
iterative& 0.2(+-0.004)	& 0.5(+-  0)	& 0.208(+-0.004)	& 0.14(+-0.004)	\\
{\bf{ approx. coevol}}& 0.511(+-0.005)	& 0.792(+-0.004)	& 0.5(+-  0)	& 0.502(+-0.005)	\\
real coevol& 0.502(+-0.005)	& 0.86(+-0.004)	& 0.498(+-0.005)	& 0.5(+-  0)	\\
\hline
\multicolumn{5}{|c|}{GuessWho (5 params)}\\
\hline
 & vs naive & vs itera. & vs ap. coevol & vs coevol \\
\hline
naive& 0.5(+-  0)	& 0.276(+-0.005)	& 0.159(+-0.004)	& 0.34(+-0.005)	\\
iterative& 0.724(+-0.005)	& 0.5(+-  0)	& 0.424(+-0.005)	& 0.583(+-0.005)	\\
{\bf{ approx. coevol}}& 0.841(+-0.004)	& 0.576(+-0.005)	& 0.5(+-  0)	& 0.563(+-0.005)	\\
real coevol& 0.66(+-0.005)	& 0.417(+-0.005)	& 0.437(+-0.005)	& 0.5(+-  0)	\\
\hline
\multicolumn{5}{|c|}{Phantom-4-in-a-row (2 params)}\\
\hline
 & vs naive & vs itera. & vs ap. coevol & vs coevol \\
\hline
{\bf{ naive}}& 0.5(+-  0)	& 0.5(+-0.005)	&   1(+-  0)	& 0.75(+-0.004)	\\
iterative& 0.5(+-0.005)	& 0.5(+-  0)	&   1(+-  0)	&   0(+-  0)	\\
approx. coevol&   0(+-  0)	&   0(+-  0)	& 0.5(+-  0)	& 0.5(+-0.005)	\\
real coevol& 0.25(+-0.004)	&   1(+-  0)	& 0.5(+-0.005)	& 0.5(+-  0)	\\
\hline
\multicolumn{5}{|c|}{Nim (383 params)}\\
\hline
 & vs naive & vs itera. & vs ap. coevol & vs coevol \\
\hline
naive& 0.5(+-  0)	& 0.5(+-0.005)	&   0(+-  0)	& 0.5(+-0.005)	\\
iterative& 0.5(+-0.005)	& 0.5(+-  0)	&   0(+-  0)	&   0(+-  0)	\\
approx. coevol&   1(+-  0)	&   1(+-  0)	& 0.5(+-  0)	&   0(+-  0)	\\
{\bf{ real coevol}}& 0.5(+-0.005)	&   1(+-  0)	&   1(+-  0)	& 0.5(+-  0)	\\
\hline
\multicolumn{5}{|c|}{Phantom-tic-tac-toe (18 params)}\\
\hline
 & vs naive & vs itera. & vs ap. coevol & vs coevol \\
\hline
{\bf{ naive}}& 0.5(+-  0)	& 0.75(+-0.004)	& 0.5(+-0.005)	& 0.5(+-0.005)	\\
iterative& 0.25(+-0.004)	& 0.5(+-  0)	&   0(+-  0)	& 0.5(+-0.005)	\\
approx. coevol& 0.5(+-0.005)	&   1(+-  0)	& 0.5(+-  0)	&   0(+-  0)	\\
real coevol& 0.5(+-0.005)	& 0.5(+-0.005)	&   1(+-  0)	& 0.5(+-  0)	\\
\hline
\multicolumn{5}{|c|}{Morra (66 params)}\\
\hline
 & vs naive & vs itera. & vs ap. coevol & vs coevol \\
\hline
naive& 0.5(+-  0)	& 0.472(+-0.005)	& 0.472(+-0.005)	& 0.477(+-0.005)	\\
iterative& 0.528(+-0.005)	& 0.5(+-  0)	& 0.475(+-0.005)	& 0.483(+-0.005)	\\
approx. coevol& 0.528(+-0.005)	& 0.525(+-0.005)	& 0.5(+-  0)	& 0.482(+-0.005)	\\
{\bf{ real coevol}}& 0.523(+-0.005)	& 0.517(+-0.005)	& 0.518(+-0.005)	& 0.5(+-  0)	\\
\hline
\multicolumn{5}{|c|}{4-in-a-row (14 params)}\\
\hline
 & vs naive & vs itera. & vs ap. coevol & vs coevol \\
\hline
naive& 0.5(+-  0)	& 0.427(+-0.05)	& 0.494(+-0.06)	& 0.503(+-0.06)	\\
iterative& 0.573(+-0.05)	& 0.5(+-  0)	& 0.494(+-0.06)	& 0.57(+-0.06)	\\
{\bf{ approx. coevol}}& 0.506(+-0.06)	& 0.506(+-0.06)	& 0.5(+-  0)	& 0.604(+-0.05)	\\
real coevol& 0.497(+-0.06)	& 0.43(+-0.06)	& 0.396(+-0.05)	& 0.5(+-  0)	\\
\hline
\end{tabular}
}
\end{subtable}%
\begin{subtable}{0.5\linewidth}
\scriptsize
{\setlength{\tabcolsep}{0.3em}
\begin{tabular}{|c|c|c|c|c|}
\hline
\multicolumn{5}{|c|}{Flip (420 params)}\\
\hline
 & vs naive & vs itera. & vs ap. coevol & vs coevol \\
\hline
naive& 0.5(+-  0)	& 0.428(+-0.005)	& 0.439(+-0.005)	& 0.439(+-0.005)	\\
iterative& 0.572(+-0.005)	& 0.5(+-  0)	& 0.437(+-0.005)	& 0.432(+-0.005)	\\
approx. coevol& 0.561(+-0.005)	& 0.563(+-0.005)	& 0.5(+-  0)	& 0.437(+-0.005)	\\
{\bf{ real coevol}}& 0.561(+-0.005)	& 0.568(+-0.005)	& 0.563(+-0.005)	& 0.5(+-  0)	\\
\hline
\multicolumn{5}{|c|}{Pig (1 params)}\\
\hline
 & vs naive & vs itera. & vs ap. coevol & vs coevol \\
\hline
naive& 0.5(+-  0)	& 0.496(+-0.005)	& 0.502(+-0.005)	& 0.505(+-0.005)	\\
iterative& 0.504(+-0.005)	& 0.5(+-  0)	& 0.506(+-0.005)	& 0.498(+-0.005)	\\
approx. coevol& 0.498(+-0.005)	& 0.494(+-0.005)	& 0.5(+-  0)	& 0.502(+-0.005)	\\
real coevol& 0.495(+-0.005)	& 0.502(+-0.005)	& 0.498(+-0.005)	& 0.5(+-  0)	\\
\hline
\multicolumn{5}{|c|}{Batawaf (3 params)}\\
\hline
 & vs naive & vs itera. & vs ap. coevol & vs coevol \\
\hline
naive& 0.5(+-  0)	& 0.49(+-0.005)	& 0.501(+-0.005)	& 0.701(+-0.005)	\\
iterative& 0.51(+-0.005)	& 0.5(+-  0)	& 0.497(+-0.005)	& 0.698(+-0.005)	\\
approx. coevol& 0.499(+-0.005)	& 0.503(+-0.005)	& 0.5(+-  0)	& 0.693(+-0.005)	\\
real coevol& 0.299(+-0.005)	& 0.302(+-0.005)	& 0.307(+-0.005)	& 0.5(+-  0)	\\
\hline
\multicolumn{5}{|c|}{War (3 params)}\\
\hline
 & vs naive & vs itera. & vs ap. coevol & vs coevol \\
\hline
naive& 0.5(+-  0)	& 0.453(+-0.006)	& 0.495(+-0.006)	& 0.508(+-0.006)	\\
{\bf{ iterative}}& 0.547(+-0.006)	& 0.5(+-  0)	& 0.563(+-0.006)	& 0.708(+-0.005)	\\
approx. coevol& 0.505(+-0.006)	& 0.437(+-0.006)	& 0.5(+-  0)	& 0.492(+-0.006)	\\
real coevol& 0.492(+-0.006)	& 0.292(+-0.005)	& 0.508(+-0.006)	& 0.5(+-  0)	\\
\hline
\multicolumn{5}{|c|}{Batawaf4 (4 params)}\\
\hline
 & vs naive & vs itera. & vs ap. coevol & vs coevol \\
\hline
naive& 0.5(+-  0)	& 0.311(+-0.005)	& 0.301(+-0.005)	& 0.385(+-0.005)	\\
{\bf{ iterative}}& 0.689(+-0.005)	& 0.5(+-  0)	& 0.507(+-0.005)	& 0.562(+-0.005)	\\
approx. coevol& 0.699(+-0.005)	& 0.493(+-0.005)	& 0.5(+-  0)	& 0.573(+-0.005)	\\
real coevol& 0.615(+-0.005)	& 0.438(+-0.005)	& 0.427(+-0.005)	& 0.5(+-  0)	\\
\hline
\multicolumn{5}{|c|}{War4 (4 params)}\\
\hline
 & vs naive & vs itera. & vs ap. coevol & vs coevol \\
\hline
naive& 0.5(+-  0)	& 0.489(+-0.008)	& 0.495(+-0.008)	& 0.493(+-0.008)	\\
iterative& 0.511(+-0.008)	& 0.5(+-  0)	& 0.424(+-0.008)	& 0.47(+-0.008)	\\
{\bf{ approx. coevol}}& 0.505(+-0.008)	& 0.576(+-0.008)	& 0.5(+-  0)	& 0.576(+-0.008)	\\
real coevol& 0.507(+-0.008)	& 0.53(+-0.008)	& 0.424(+-0.008)	& 0.5(+-  0)	\\
\hline
\multicolumn{5}{|c|}{Cheat-Special (29 params)}\\
\hline
 & vs naive & vs itera. & vs ap. coevol & vs coevol \\
\hline
naive& 0.5(+-  0)	& 0.497(+-0.01)	& 0.499(+-0.01)	& 0.5(2\textsuperscript{th} +-0.01)	\\
{\bf{ iterative}}& 0.503(+-0.01)	& 0.5(+-  0)	& 0.658(+-0.01)	& 0.5(+-0.01)	\\
approx. coevol& 0.501(+-0.01)	& 0.342(+-0.01)	& 0.5(+-  0)	& 0.5(+-0.01)	\\
real coevol& 0.5(+-0.01)	& 0.5(+-0.01)	& 0.5(+-0.01)	& 0.5(+-  0)	\\
\hline
\end{tabular}
}
\end{subtable}
\caption{\label{popof}\small Comparison in terms of winning rate between the naive and iterative approaches and coevolution. The value after `$\pm$' is the standard deviation. The best method (if any) if the one which wins with probability at least $50\%$ against all others. It is indicated in bold. It is possible that more than one method wins with probability at least $50\%$ against all others (e.g., for ``Phantom-tic-tac-toc", it is the case for naive and real coevolution, however, naive is better). The coevolution methods dominate in most case (in $8$ games over $13$).}
\end{table*}

%\subsubsection{Guess who}
%In the Guess who game, both players have a set of cards (e.g. 24 cards).%\footnote{It is often played with only 24 cards.}. 
%An unknown card is distinguished, for both players - it is randomly drawn, and it is not necessarily the same for both players. Each player, in turn, selects a subset of his cards; either his subset, or the complementary subset, depending on which one contains his/her distinguished card, is preserved while the rest is removed. The first player with only one card remaining (this remaining card is necessarily the preserved card) has won.

\def\toolong{{\color{magenta} might be too long; maybe the short version commented in the tex is enough...\\

``Guess who ?''  is a two-player game which appeared in 1979 in Great Britain (\url{en.wikipedia.org/wiki/Guess_Who}). 

Each player has a set of characters, plus a hidden character which belongs to the opponent's set of characters. Each player, in turn, asks a ``yes/no'' question about the opponent's character, and the opponent answers this question; and then the asker keeps only in his set the characters who are consistent with the answer. The winner is the first player whose set of characters has cardinal one. 

We here assume that there is no error, i.e. the set of characters always contains the opponent's hidden character. We also assume that we are not allowed to hide our current set of characters (which could theoretically be kept in memory, thanks to the sequence of questions).

The game is finite, so there exists an optimal policy in spite of the partial observability. This optimal policy must ensure (when the first player is randomized) at least 50 \% winning rate against any opponent. Indeed, using the symmetries of the game, the game boils down to a random fully observable game. As shown by several results in the present paper, the optimal policy is however not the one which reaches the best winning rate against the (classical, defined below) dichotomy policy.
}}

\def\nobodycares{
\subsubsection{Properties of these games}
Table \ref{properties} summarizes the properties of the considered games.
\begin{table}[h]
\center
\begin{tabular}{|c|c|c|c|c|}
\hline
Name of          & Stochastic & Partially & Can be made      & Necessarily\\
the		 & transition & observable& fully observable & stochastic\\
game             &            & or simul. & and turn-based   & mixed Nash\\
		 &	      &	actions	  & by introducing   & policies\\
		 &		&	  & stochastic       &   \\
		 &		&	  & transitions	     &   \\
\hline
Flip		 & No		& Yes	  & Yes              & No \\
Morra 		 & No		& Yes     & No               & Yes \\
Battleship       & No           & Yes     & No               & Yes \\
Cheat		 & No		& Yes	  & No		     & ?\\
%Nim		 & No		& No	  & ---		     & No \\
Phantom & No		& Yes     & No               &  ? TODO2\textsuperscript{th}  \\
tic-tac-toe & & & & \\
Guess who ?        & No		& Yes     & Yes		     & No \\
Phantom  & &  & & \\
4-in-a-row & No & Yes & No & \\
Batawaf & No & Yes & No& ?\\
War & No & Yes & No& ?\\
\hline
\end{tabular}
\caption{\label{properties}TODO Except Nim, all these games are deeply stochastic ().}
\end{table}
TODO keep battleship ?
}

\section{Detailed strategies}\label{sec:detailed}

This section studies some specific games, and shows that our parametric strategies could find out some surprisingly strong strategies against the baselines. Some of these strategies can make you really strong at Guess who (which was already known, but we provide human readable strategies) and at Batawaf or War (which is more unexpected). %{\color{magenta}Policies at Cheat, Battleship and Flip are less readable.}
%Other results are presented in Section \ref{others}

\subsection{Batawaf and game of War: good cards first!}\label{bw}

The rules do not specify in which order the cards won by a player should be added at the end of the deck.
This order has an impact on the length of games, and on the probability that the game finishes: loops are possible, see~\cite{funnyguys}. We here stop the game and consider it as a draw after $1$ million moves without conclusion.

\paragraph*{Considered strategies}
We consider 4 parameters, namely $A,B,C,D$, which are converted into non-negative parameters using $a=\exp(A)$, $b=\exp(B)$, $c=\exp(C)$, $d=\exp(D)$.
Then, when we must put $k$ won cards in our deck (at the end), their order is chosen as follows:
\begin{enumerate}
\item $d$ is used as a seed for generating a permutation $\pi$ of $\{1,2,\dots,k\}$;
\item $\sigma$ is a permutation of $\{1,2,\dots,k\}$:
	\begin{itemize}
		\item uniformly drawn with probability $a/(a+b+c)$;
		\item equal to $\{1,2,\dots,k\}$ with probability $b/(a+b+c)$;
		\item equal to $\{k,k-1,\dots,1\}$ with probability $c/(a+b+c)$;
	\end{itemize}
\item the cards are put in ascending order and shuffled using permutation $\pi\circ \sigma$.
\end{enumerate}

\paragraph*{ Results \& discussion } For some specific values, we therefore get the following human-readable strategy: put the cards at the end of your deck in decreasing order, the best card first. This strategy is termed ``descending'' in the rest of this paper. The naive strategy consists in randomly sorting the card. The ascending strategy is the opposite of the descending one.
An alternate version with 3 parameters uses only $A,B,C$ and the identity for $\pi$. Its optimization was much faster, immediately converging to the use of descending order, our best strategy so far.

In batawaf, the descending strategy gets winning rate 68.8\% against the naive strategy, 57.5\% against the ascending strategy, and $50\%$ against itself (tested on 10000 games).
In War, the corresponding performances are $70.0\%$, $53.3\%$ and $50\%$.

\subsection{Guess who: play risky moves when late!}\label{gw}

%{\color{magenta} too many things here, shorter version needed\\

In many cases, we have an intuitive understanding of what is a ``risky'' move, and more precisely a ``good risky move''. It is a move which, simultaneously, maximizes the expected long term reward, and has a large probability of making, at a tactical level, the situation worse than if we had played a simple move. %We here investigate this concept in the case of the ``Guess who ?'' game. Somehow surprisingly, we design strategies much better than the classical dichotomy policy.

%When choosing an option in a finite set, a classical framework is the bandit one. In such a setting, various concepts have been tested\cite{nipsrisk,nipsrisk2,bubeckrisk}; but the goal is to get rid of risk, not to make relevant risky decisions. In industry or agriculture\cite{shierisk,YAN}, risk is also considered as a problem; it is also an enemy in many works around finance\cite{riskrl} and risk-averse behaviors can arise in bargaining games\cite{bargrisk} or variants of matching pennies\cite{riskmp}. However in economy risk-seeking is also considered, in particular for oil exploration\cite{oilrisk}, film making, and more generally domains in which one big success makes dozens of failures irrelevant. In this last case, we see a dilemma between tactics and strategy; we make a decision likely to fail, because the small probability of great success is indeed critical at the strategic level.

%There are psychology studies around risk. In terms of motivation, risk-seeking is attributed to age\cite{adorisk}, gender\cite{genderrisk}, and attributed to long term rewards (including both humans and animals\cite{riskhumansanimals}). In Chess, risk-taking has been correlated to culture\cite{chessrisk} or even ``facial masculinity''\cite{facialrisk} and the beauty of their (female) opponent (for males only) \cite{beautyrisk}. Maybe teenagers or male players in front of women have a strategic advantage in playing risky; this is beyond the present work.

In games, choosing the right level of risk depends on the current situation. If the goal is to win, then, in a territory game such as Go, ensuring an almost sufficient territory is meaningless; it is better to have a 5\% probability of having more territory than your opponent, rather than a probability 100\% of having just a little bit less territory than your opponent. Therefore, the ``shobute'' (\url{http://senseis.xmp.net/?Shobute}) principle states that, if you are behind in a game of Go, it is time for a risky attack.
%In martial arts, a related concept is ``sutemi''. For example, in Judo, if you are going to lose equilibrium anyway, do not try to keep your equilibrium - the tactically risky decision of losing equilibrium but using this loss of equilibrium for attacking (typically, falling on one's back and throwing the opponent) is strategically less risky than trying to keep equilibrium.
Another case occurs when playing a sequence of games, with, e.g., half a point in case of draw, one point in case of win, and zero in case of loss. Then, the player in advance might prefer to play very safely, ensuring a draw, whereas the other must play risky.
%%%%In terms of artificial intelligence, it has been said that MCTS\cite{coulom06,mogofpu,uct}
%%%% has this property that it plays risky agressive moves when it is late, and that it is not 
%%%%too bad; but also that, when it is in advance, it can be play stupidly
Here, we investigate the case of Guess who and show that, contrarily to what is usually assumed, the optimal strategy must consider the necessity of taking risks.
%}

Let us discuss the Guess who game. First, for any arbitrary subset of your characters, you can design a question which discriminates exactly this subset.  This can be done by using long questions (``is your character Tom or Jean or Pierre or Chang-Shing?''), or the alphabetic order. Randomizing the permutation, we get a strategy which is invariant by permutation of the characters. As a consequence, the order and traits do not matter. We can therefore work in a permutation-invariant manner, and just split between the $c$ first and the $n-c$ last characters, when we have a list of $n$ characters. $c<1$ or $c>n-1$ do not make any sense; therefore we have $n-2$ possible actions (up to symmetries) when we have $n$ characters.  Using these symmetries, the state of our set of characters become simply a number between $1$ and $N$, where $N$ is the original number of characters. This state is just the number of remaining characters. We can use the number of characters in the opponent's set as well, as a side information; we might want to make more risky decisions (farther from equal split) when we have a bad situation. This is precisely the point in this paper.  As a consequence, up to symmetries (permutations of our characters and of the opponent's characters) our strategy uses as an input the pair $(n,m)$ ($n$ is our number of characters and $m$ is the number of characters of our opponent) and outputs $1<c<n-2$.

A simple first remark is that dichotomy (constructing a question such that approximately half characters correspond to ``yes'') is a reasonably good strategy. %, recommended in {\small{\url{blog.teamtreehouse.com/binary-search-or-how-to-win-at-guess-who}}}. As pointed out in this blog, the traits are chosen so that such questions are difficult to find, but, for example using alphabetic order, we can implement this policy.  The same policy is proposed in many references {\small{\url{www.cracked.com/article_21940_5-mathematical-strategies-dominating-popular-kids-games.html}, \url{sciencemagician.wordpress.com/2014/10/17/guess-who-winning-strategies/}, \url{stackoverflow.com/questions/26102473/guess-who-best-algorithm}, and in \url{boards.straightdope.com/sdmb/archive/index.php/t-627593.html}.}}
%\subsubparagraph{Advanced strategies}
As pointed out in 
{\small\url{boardgamegeek.com/thread/302791/advanced-strategies}},
when you are late, you should not apply the dichotomy otherwise you are just going to lose. Instead, you should use risky strategies, with which you might win early.
This reference also considers hiding the current size of the set of characters. They however do not provide any experiment or formalization of strategies.
%\section{Strategies}

\paragraph*{Strategies}
The dichotomy strategy is simply:
 $$(n,m) \mapsto \lfloor n/2 \rfloor.$$
%The randomized policy is not very efficient; it is $(n,n’) \mapsto \lfloor rand \times n/2 \rfloor$.
%The following policy makes more risky choices (at least if $\alpha>0$) when the opponent has a small set of characters (i.e. when the opponent has a good situation): $(n,n’) \mapsto \max(1, \lfloor n/2 - \alpha \max (n-n',0)/2 \rfloor)$.
A general strategy, which covers the dichotomy case, is:
\begin{scriptsize}
\begin{equation}
\pi_{\alpha,\beta}(n,m)=\max\left(1, \left\lfloor \beta r\frac{n-1}{2} + (1-\beta) \left(\frac{n}{2}-\alpha\frac{\max(n-m,0)}{2} \right) \right\rfloor \right),\label{lineareq}
\end{equation}
\end{scriptsize} 
%this covers all previous policies. 
with $\alpha\in\R$, $\beta\in\R$ and $r\in[0,1]$ drawn uniformly at random.
It will turn out, however, that $\beta>0$ is a poor strategy. We can further extend it by including non-linear terms:
\begin{scriptsize}
\begin{equation}
\pi_{\alpha,\beta,\gamma}(n,m)=\max\left(1, \left\lfloor \beta r \frac{n-1}{2} + (1-\beta) \left(  \frac{n}{2}-\alpha\frac{\delta_m}{2} + \gamma \frac n2 \frac{\delta_m^2}{n^2} \right) \right\rfloor \right),\label{nonlineareq}
\end{equation}
\end{scriptsize}
where $\delta_m=\max(n-m,0)/2$, $\alpha\in\R$, $\beta\in\R$, $\gamma\in\R$ and $r$ as in Eq.~\ref{lineareq}.

We can consider yet another formula, using $\beta=0$ and introducing two new parameters $\zeta\in\R$ and $\iota\in\R$, as follows:
\begin{scriptsize}
\begin{equation}
\pi_{\alpha,\gamma,\zeta,\iota}(n,m)=\max\left(1, \left\lfloor  \frac n2-\alpha\frac{\delta_m}2 + \gamma \frac n2\frac{\delta_m^2}{n^2} +\zeta \frac n2\frac{\delta_m^3}{n^3}+\iota \frac{n-m}2 \right\rfloor \right),\label{nonlineareq2}
\end{equation}
\end{scriptsize}
with $\alpha\in\R$, $\gamma\in\R$ and $\delta_m$ as in Eq.~\ref{nonlineareq}.
%   d=floor(b*rand*(n-1/2) +(1-b)*(n/2-a*max(n-np,0)/2) +(1-b)*c*(n/2-((max( n-np,0)/(n))^2)*(n/2)) );
%(1-b) gamma((n/2-ret)/n)^2\times n/2
%Some specific policies will arise in the present work.
%Dichotomy is the baseline discussed above; it is often assumed that it is optimal, and the present work shows that this is not the case, by far.

\paragraph*{Results and discussion}
Strategies of Eqs.~\ref{lineareq}, \ref{nonlineareq} and \ref{nonlineareq2} are first optimized in terms of winning rate against the dichotomy strategy (4 hours of optimization). It is a simple application of the naive approach. Since it appeared that $\beta>0$ in Eqs.~\ref{lineareq} and \ref{nonlineareq} leads to poor strategies, we dismissed this term (i.e. $\beta=0$). At this step, the optimal strategy then corresponds to $\alpha=1$ and $\beta=0$ in Eq. \ref{lineareq}, or equivalently , $\alpha=1$ and $\beta=0$ and $\gamma=0$ in Eq. \ref{nonlineareq}, or equivalently $\alpha=1$ and $\gamma=\zeta=\iota=0$ in Eq. \ref{nonlineareq2}. We call this strategy \textit{optimal linear}, shortened in OL.

Second, applying the iterative approach, we optimize a family of non-linear strategies ( Eq.~\ref{nonlineareq2}) by maximizing the winning rate against the OL strategy, using Eq.~\ref{nonlineareq} under constraint $\beta=0$ (4 hours of optimization). The optimal parameters are $\alpha=-\frac14$ and $\gamma=-\frac32$, $\zeta=\iota=0$. This strategy is termed \textit{optimal nonlinear}, denoted ONL.

Finally, in one more step of the iterative approach, a better 4-parameter strategy, termed `̀ Best'', optimal strategy against ONL (4 hours of optimization), is obtained with $\alpha=-0.56$, $\gamma=-1.58$, $\zeta=-0.06$, $\iota=-0.022$ in Eq.~\ref{nonlineareq2}.

\begin{figure*}
\center
\includegraphics[width=.45\textwidth]{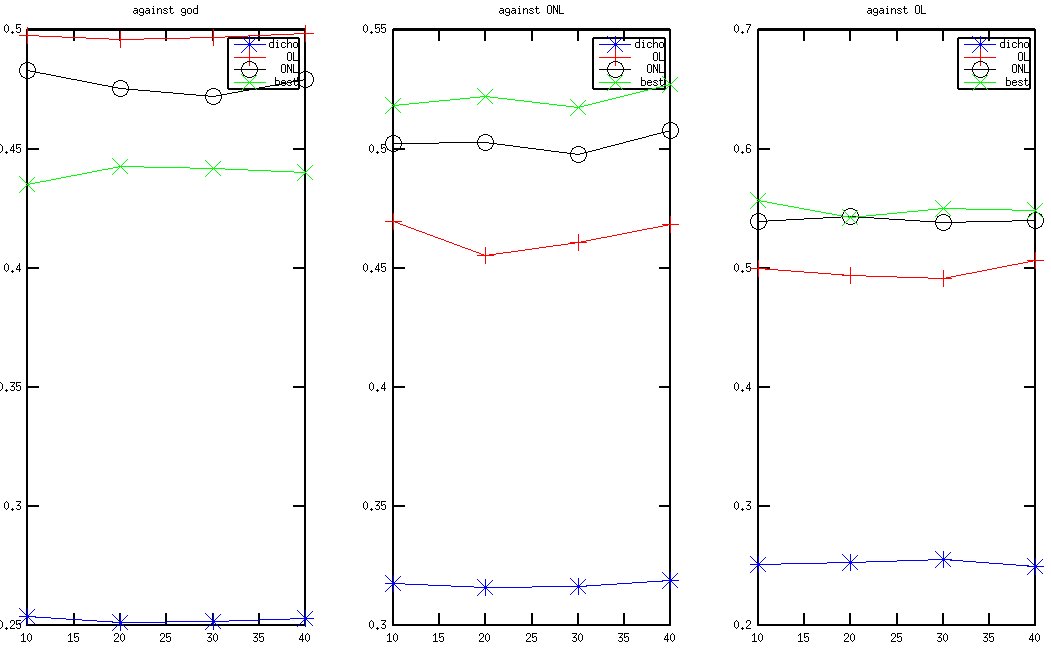}
\caption{\label{fig:gw}\small Results at Guess who (winning rate on the y-axis, as a function of the initial number of characters on the x-axis). We present results of OL, ONL, Best and the simple dichotomy (which performs very poorly). Left: performance against the optimal strategy (termed ``God'') designed in \cite{guesswho}. Middle: performance against ONL. Right: performance against Dichotomy. Essentially the OL method performs well against God, though it could be slightly exploited by a method optimized against it. Interestingly methods optimized against OL were on the other hand weaker against God.}
\end{figure*}

\cite{guesswho} pointed out the existence of an analytical optimal strategy: it is used in our tests. Results are presented in Fig. \ref{fig:gw}. We see (right) that ONL wins against OL, as well as Best. Best wins against ONL (middle), but against God (the optimal strategy) only OL reaches 50\%. The naive approach can be ``exploited'' (in the Nash sense: a strategy specialized against it will outperform it) but it is quite strong both against the baseline and against God (approximately 50\%: see left).

% We see that the naive approach performs quite well; whereas iterated optima suffer from a red queen effect and become weak against the optimal policy.
% OL (optimal linear) corresponds to $\alpha=1$ and $\gamma=\zeta=\iota=0$ in Eq. \ref{nonlineareq2}, or equivalently $\alpha=1$ and $\beta=0$ in Eq. \ref{lineareq}, or equivalently $\alpha=1$ and $\beta=0$ and $\gamma=0$ in Eq. \ref{nonlineareq}. This policy arises in the optimization of Eq. \ref{lineareq} or of Eq. \ref{nonlineareq}, after the term in $\beta$ has been dismissed (i.e. $\beta=0$); they are optimized in terms of winning rate against the dichotomy policy.
% ONL (optimal nonlinear) corresponds to $\alpha=-\frac14$ and $\gamma=-\frac32$, $\zeta=\iota=0$. It arises in Fig. \ref{nonlinearres} (bottom), i.e. when optimizing the winning rate against OL, using the policy described by Eq. \ref{nonlineareq} under constraint $\beta=0$ and in the context of 128 characters. It is shown in Fig. \ref{boplot}
% A better 4-parameter policy, termed `̀ Best'', outperforming ONL, is provided in Fig. \ref{opof}; it uses $\alpha=-0.56$, $\gamma=-1.58$, $\zeta=-0.06$, $\iota=-0.022$ in Eq. \ref{nonlineareq2}.
%A very recent paper \cite{guesswho} pointed out the existence of an analytical optimal policy; it is used in our tests. Results are presented in Fig. \ref{gw}. We see that the naive approach performs quite well; whereas iterated optima suffer from a red queen effect and become weak against the optimal policy.

\subsection{Cheat game: ensuring some draw}
At the cheat game, both the naive and iterative methods found a strategy with a lot of draws.
Observing games, we could extract the following optimal strategy, ensuring a draw:
\begin{enumerate}
	\item If your opponent has more than 3 cards, play an arbitrary strategy.
	\item If your opponent has 2 or 3 cards, and it is your turn to choose a current color, then put a card of a color that he does not have, and say ``cheat'' when he plays; he has now more than 2 cards.
	\item If your opponent has 3 cards and it is your turn to put a card (the current color already being set), then say ``cheat''. Either he has cheated and has now more than 3 cards, or not and you are now in the situation of point 2.
\end{enumerate}
This strategy ensures that the opponent has never less than 2 cards.

\section{Conclusion}
% We provide experimental results for parametric policies, mainly for partially observable games.
\subsection{Comparing algorithms}
For many games, simulating the current hidden state conditionally to partial observation is tricky, and mathematically principled algorithms for this\footnote{Keep in mind the undecidability result in \cite{auger2012frontier}, which shows that a general algorithm for belief state information cannot exist in the general case of partially observable game.} are hardly available. For such games, implementing a parametric strategy and optimizing the parameters (possibly with zero human knowledge) works: we obtained new interesting strategies in Cheat (never-losing strategy, found before humans solved it), Guess who (on par with mathematically exact method), War \& Batawaf (strategy choosing the order of won cards put at the end of one's desk). In phantom cases and some others we just optimize the complete deterministic strategy, i.e., we use zero human knowledge. Inside this general principle of little or zero expertise parametric policy optimization, comparing algorithms is difficult; the conclusions are never clear and universal. We cannot claim a strong superiority of any of our methods against other ones. We might advocate a portfolio of methods, i.e., testing several methods.

For partially observable games, rankings between players are not clear. The Elo model~\cite{elo} does not apply because superiority at a given game is not transitive (see the 4 strategies obtained for phantom-tic-tac-toe by the naive method, the iterative method, coevolution and approximate coevolution). Cycles make the naive and iterative method risky and sensitive to the red queen effect.

The iterative method is appealing compared to the naive one when the Elo model applies, because optimizing against a fixed baseline might become very hard, in particular when the success rate is already quite high, so that fitness evaluations are not very informative - a Bernstein race between candidates close to 100\% winning rate is very slow.

Results on Guess who, including the mathematical exact optimum, show that we can compete in a couple of lines of code and a few hours of computation with a mathematical analysis (see the OL method, obtained by naive optimization against the dichotomy). However, our method can be (slightly) exploited by a method optimized specifically against it, whereas the mathematical method cannot (by definition, and our experiments confirm this) be exploited (Fig. \ref{fig:gw}). 

The most surprising results for us was the successes of the seed method. This method just generates a large population, and simulates games (possibly in a more clever manner than full Round-Robin, see e.g. \cite{fastseed}) for choosing the best - coevolution, but with no mutation, no iteration, no crossover, just random generation and selection.

In particular, it does work when nothing works, e.g. in the battleship case (too expensive for other methods - we have time for only 54 games in this platform in the considered time setting!), or when the strategy is very poorly parametrized so that even a trivial game becomes complicated (e.g. Nim). It was never far from the best, except when it was possible to solve exactly the game with 100\% winning rate against the baseline as in 4-in-a-row.

\subsection{Results on specific games}
In Batawaf and the card game of War, we provide a surprisingly strong heuristic. To the best of our knowledge, nobody ever published such strong strategies at these games, while War is one of the most widely played games in the world. We get close to 70\% winning rate against the baseline that most people play. 
In Battleship, we provide both a probability distribution for the initial positioning of the fleet and for the Monte-Carlo simulation of the opponent's fleet. Only the seed method could find something meaningful.
In Guess who, we provide a strategy which is on par with the mathematically derived optimum, which is the key output of a paper published as a mathematical study. The algorithm derived, by itself, an application of the ``shobute'' principle, namely we should have more risky decisions when we are late. We ran our experiments before the mathematical solution was published.
At Cheat, we get a strong strategy, including against humans. This was found by computers before being understood by humans.
%choose a game with high stochasticity and hard to remember everything ⇒ there are good “positional” strategies ⇒ parametric policies on top of features
%build very stochastic opponent
%optimize against this very stochastic opponent (TODO which oponents are highly stoc ?)
%batawaf, war: maximize wr

%\bibliographystyle{abbrv}
\bibliographystyle{IEEEtran}
\bibliography{ref}
\end{document}